\documentclass[aps,8pt,prd,twocolumn,preprintnumbers,amsmath,amssymb,nofootinbib,a4paper]{revtex4}
\pdfoutput=1
\usepackage[colorlinks=true, pdfstartview=FitV, linkcolor=blue, citecolor=red, urlcolor=magenta]{hyperref}

\usepackage{graphicx}
\usepackage{microtype}
\usepackage{amsmath}
\usepackage{amssymb}
\usepackage{subfigure}
\usepackage{hyperref}
\usepackage{url}
\usepackage{xcolor}
\usepackage{color}
\usepackage{mathrsfs}
\usepackage{calrsfs}
\usepackage{amsfonts}
\usepackage{eufrak}
\usepackage{tabularx}
\usepackage{eucal}
\usepackage{latexsym}
\usepackage{ragged2e}
\usepackage{epsfig}
\usepackage{textcomp}
\usepackage{subfigure}
\usepackage{marvosym}
\usepackage{ifsym}
\usepackage{lipsum}
\usepackage{appendix}
\usepackage{bm}
\usepackage{mathrsfs}

 \hypersetup{
 	colorlinks=true,
 	linkcolor=red,
 	citecolor=blue,
    urlcolor=amaranth,
 }
\definecolor{vividviolet}{rgb}{0.62, 0.0, 1.0}
\definecolor{amaranth}{rgb}{0.9, 0.17, 0.31}
\definecolor{palatinateblue}{rgb}{0.15, 0.23, 0.89}
\definecolor{brightpink}{rgb}{1.0, 0.0, 0.5}
\definecolor{cornflowerblue}{rgb}{0.39, 0.58, 0.93}
\definecolor{deepcarminepink}{rgb}{0.94, 0.19, 0.22}
\definecolor{radicalred}{rgb}{1.0, 0.21, 0.37}
\hypersetup{linktoc=all,
    colorlinks={true},linkcolor={palatinateblue},
    citecolor={brightpink}, urlcolor={amaranth}}

\begin{document}

\thispagestyle{empty}

\begin{center}

\title{Insight into the Microstructures of Black Holes with Quantum-Triggered Violations}

\author{Haximjan Abdusattar}
\email{axim@ksu.edu.cn}
\affiliation{School of Physics and Electrical Engineering, Kashi University, Kashgar 844000, Xinjiang, China}

\begin{abstract}

In this study, we investigate the microstructures of a charged AdS (Anti-de Sitter) black hole exhibiting quantum anomalies through the lens of Ruppeiner geometry. Previous research has established that black holes undergo $P$-$V$ phase transitions and exhibit critical phenomena near their critical points, characterized by four critical exponents that typically obey scaling laws predicted by mean-field theory. However, recent findings have revealed that black holes with quantum anomalies can violate these scaling laws. Motivated by these discoveries, we employ Ruppeiner geometry to probe the thermodynamic fluctuations and gain insights into the microstructure of such black holes. Our analysis aims to elucidate how quantum effects modify the microscopic properties of spacetime, offering a novel perspective on the understanding of black hole thermodynamics.

\end{abstract}

\maketitle

\end{center}

\section{Introduction}

Since the pioneering work that established the four laws of black hole thermodynamics \cite{Bardeen:1973gs}, the field of black hole physics has evolved significantly. The identification of surface gravity and area as black hole temperature \cite{Hawking:1975vcx} and entropy \cite{Bekenstein:1973ur}, respectively, has opened up new avenues of research in black hole thermodynamics. One of the most vibrant research areas in this domain centers on exploring phase transitions in black holes, particularly within the fascinating realm of black hole chemistry in the extended phase space framework, where the cosmological constant is viewed as a thermodynamic pressure variable, $i.e.$, $P=-\Lambda/8\pi$ \cite{Kastor:2009wy,Dolan:2010ha,Dolan:2011}.
With such an interpretation, in Ref.\cite{Kubiznak:2012wp} the authors has demonstrated that four-dimensional charged AdS (Anti-de Sitter) black holes exhibit both first-order (discontinuous) and second-order (continuous) phase transitions between small and large black hole states within this extended phase space. This transition mirrors the van der Waals gas/liquid phase transition, resulting in identical critical exponents for both systems, and hence obey the thermodynamic scaling laws as described by mean field theory.
The study of critical behavior in black holes within this context, known as ``$P$-$V$ criticality" has been extensively explored in the literature \cite{Gunasekaran:2012dq,Wei:2012ui,Cai:2013qga,Dehghani:2014caa,Hendi:2012um} (see also \cite{Xu:2015rfa,Cai:2014znn,Majhi:2016txt,
Cheng:2016bpx,Dehyadegari:2017hvd,Kumara:2019xgt,Li:2020xkh,Hu:2018qsy,Bhattacharya:2017hfj,Bhattacharya:2017nru,Hendi:2017fxp,Abdusattar:2025rdp} for more related works and reviews \cite{Altamirano:2014tva,Kubiznak:2016qmn}). Various intriguing phenomena have been observed in the extended phase space of black holes, including multiple critical points \cite{Tavakoli:2022kmo}, triple critical points \cite{Altamirano:2013uqa}, reentrant phase transitions \cite{Altamirano:2013ane}, and superfluid-like phase transitions \cite{Hennigar:2016xwd}.

Despite the significant progress in black hole thermodynamics, the statistical description of black hole microstates remains elusive. The lack of a complete understanding of the microscopic structure of black holes has profound implications for our comprehension of quantum gravity. To address this issue, various approaches and theories of gravity have been explored \cite{Geng:2024jmm,Maldacena:1996gb,Horowitz:1996fn,Emparan:2006it,Lunin:2002qf,Lu:2009em,Ashtekar:1997yu}, with the aim of discerning the underlying degrees of freedom of a black hole.
One intriguing way to incorporate quantum effects in black holes is to consider the back-reaction caused by conformal anomaly of quantum field theory in black hole spacetime \cite{Cai:2009ua,Cai:2014jea}. In classical field theory with conformal symmetry, the trace of the stress-energy tensor vanishes. However, for quantum conformal field theory (CFT) in four dimensions, its vacuum expectation value \cite{Duff:1993wm,Deser:1993yx}
\begin{eqnarray}\label{QanomalyST}
    \langle T^{\mu}_{\mu} \rangle_{a}&=&\beta C_{\mu\nu\lambda\delta}C^{\mu\nu\lambda\delta}- \nonumber\\
    &&\alpha_{c}(R_{\mu\nu\lambda \delta}R^{\mu\nu\lambda\delta}-4R_{\mu\nu}R^{\mu\nu}+R^{2})\,,\,\,\,\,\,\,\,\,
\end{eqnarray}
is non-zero, leading to the phenomenon known as quantum conformal anomaly. The quantum conformal anomaly can be expressed mathematically in terms of the Weyl tensor $C_{\mu\nu\lambda\delta}$ and the Euler characteristic (the second term of r.h.s in above equation), with two non-negative constants called central charges that reflect the degrees of freedom of the underlying quantum field theory \cite{Duff:1993wm,Deser:1993yx}. Considering the back-reaction induced by this quantum anomaly, the classical solution to the Einstein equation undergoes modification, resulting in a novel spherical symmetric black hole solution \cite{Cai:2009ua,Cai:2014jea}
\begin{equation}
    ds^{2}=-f(r)dt^{2}+{f(r)}^{-1} dr^{2}+r^{2}(d\theta^{2}+\sin^{2}\theta d\phi^{2})\,\,\,\,
\end{equation}
where the function $f(r)$ is given by\footnote{In this scenario, the branch with the negative sign ($``-"$) pertains to an AdS spacetime, whereas the branch with the positive sign ($``+"$) corresponds to a de Sitter spacetime. However, in this paper, we are solely interested in the AdS scenario and its thermodynamic properties.}
\begin{equation}\label{fr}
    f(r)=1-\frac{r^{2}}{4\alpha_{c}}\left[1\pm \sqrt{1-8\alpha_{c}\Big(\frac{2M}{r^{3}}-\frac{Q^{2}}{r^{4}}-\frac{1}{l^{2}}\Big)}\right]\,,
\end{equation}
considering only type $A$ anomaly with $\beta=0$. Here, $M$ is the mass of the black hole, $Q$ can be interpreted as the $U(1)$ charge of the underlying conformal field theory, and $\alpha_{c}$ is the central charge, while $l$ is linked to the cosmological constant via $\Lambda=-{3}/{l^{2}}$. In the limit of large $r$, where quantum corrections are negligible, this solution approximates to the Reissner-Nordstr\"{o}m-Anti-de Sitter (RN-AdS) black hole.
%
This black hole solution, which incorporates the effects of quantum conformal anomaly, exhibits interesting thermodynamic properties. By examining the thermodynamic properties of this black hole solution, a deeper insight can be gained into the black hole's microscopic structure and the fundamental degrees of freedom that govern its behavior.

Interestingly, a recent study has unveiled a fascinating discovery: in black hole systems with quantum conformal anomaly, the scaling laws can be violated under specific conditions \cite{Hu:2024ldp}. When $\alpha_{\rm c} \leqslant 0$, the black hole system aligns with mean field theory: the scaling laws (describing relationships between the system's physical quantities) remain intact, indicating quantum conformal anomaly does not significantly alter the standard behavior expected from mean field theory under these conditions. However, when $\alpha_{\rm c}={Q^{2}}/{8}$, a striking breakdown of scaling laws occurs in the black hole system. This specific condition arises when the coefficient $a_{11}$ in the expansion of the pressure $P(V,T)$ near the critical point is equal to zero. Here, the central charge parameter $\alpha_{\rm c}$ and $U(1)$ charge $Q$ exhibit a correlation that induces a unique phase structure in the black hole system. This phase structure differs significantly from that observed in van der Waals-like systems, which typically adhere to scaling laws. The violation of scaling laws in this scenario highlights that the quantum conformal anomaly plays a crucial role in modifying the behavior of the black hole system under these specific conditions.
Notably, the critical phenomenon of non-standard critical exponents was first observed in another type of black hole system \cite{Frassino:2014pha}, and subsequent work has proposed a new interpretation for the physical origin of such non-standard critical behaviors \cite{Ahmed:2022kyv}, providing valuable insights for understanding the critical behavior of quantum-corrected black holes.

Despite discovering a novel phase transition, the statistical description of black hole microstructures exhibiting quantum anomalies has not yet been thoroughly investigated, profoundly influencing our comprehension of quantum gravity. Notably, the thermodynamic geometry method emerges as a pivotal tool for probing into the microstructures of black holes. In Ref.\cite{Ruppeiner:1995zz}, Ruppeiner introduced a Riemannian thermodynamic entropy metric to delineate the theory of thermodynamic fluctuations and delineated a systematic framework for computing the Ricci curvature scalar $R$ of the Ruppeiner metric, also known as the thermodynamic curvature. This metric revealed a profound connection between the sign of $R$ and the particle interactions within a thermodynamic system: a positive $R$ signifies repulsive interactions, a negative $R$ indicates attractive interactions, and $R=0$ implies the absence of interactions \cite{Ruppeiner:2012uc,Ruppeiner:2013yca,Oshima}. Since its inception, thermodynamic geometry has been meticulously explored in the context of various black hole systems
\cite{Sahay:2010tx,Niu:2011tb,Aman:2003ug,Sarkar:2006tg,Quevedo:2008xn,Banerjee:2010bx,Wei:2019uqg,Wei:2019yvs,Wei:2015iwa,Zhang:2015ova,HosseiniMansoori:2020jrx} (for related works, see \cite{Hu:2020pmr,Abdusattar:2023xxs,Hendi:2015xya,Sahay:2016kex,Wei:2019ctz,Wei:2020poh,Dehyadegari:2016nkd,KordZangeneh:2017lgs,Xu:2019gqm,Ghosh:2019pwy,Ghosh:2020kba,Guo:2019oad}), and our work specifically focuses on the application within the FRW cosmological model \cite{Abdusattar:2023pck}.
Inspired by these fascinating discoveries, our research endeavors to utilize Ruppeiner geometry to delve into the microstructure of black holes characterized by quantum conformal anomalies. By examining the curvature of the Ruppeiner metric, we aspire to gain deeper insights into the nature of microscopic interactions and the underlying degrees of freedom that govern the behavior of these black holes.

The structure of the paper is organized as follows. In Sec. \ref{sec:Ruppeiner}, we shall make a brief review on Ruppeiner thermodynamic geometry. In Sec.\ref{Rup-BHanomaly}, we shall study the behavior of thermodynamic curvature for the charged AdS black hole with quantum anomaly with a central charge $\alpha_{\rm c} \leqslant 0$, where we obtain the coexisting volumes of the black hole during the $P$-$V$ phase transition.
In Sec. \ref{sec:PV20}, we investigate the $P$-$V$ criticality and the behavior of the thermodynamic curvature of black holes with a positive central charge ($0<\alpha_{c}<{Q^{2}}/{8}$).
In Sec.\ref{RupBHaQ}, we shall obtain the coexisting volumes of the charged AdS black hole with quantum anomaly during the $P$-$V$ phase transition by considering the specific relation $\alpha_{c}={Q^{2}}/{8}$, corresponding to the case that the black hole system exhibits a breakdown of scaling laws. We also study the behavior of thermodynamic curvature of this black hole along coexisting volumes. In Sec.\ref{conclusion}, we end the work with conclusions and discussion. Throughout this paper, we adopt natural units such that $c=G=\hbar=1$.

\section{Warm-up: Ruppeiner thermodynamic geometry}\label{sec:Ruppeiner}

In this section, we will provide a concise overview of some key results of Ruppeiner thermodynamic geometry \cite{Ruppeiner:1995zz,Ruppeiner:2012uc,Ruppeiner:2013yca,Wei:2019uqg,Wei:2019yvs} as an introduction. While statistical mechanics allows us to derive macroscopic thermodynamic quantities from the microscopic constituents of an ordinary fluid system, our current knowledge necessitates adopting a thermodynamic geometric approach in reverse to understand the microscopic structure of a black hole. We start with Boltzmann's entropy formula \cite{Ruppeiner:1995zz}
\begin{equation}
 S=k_{\rm B}\ln\Omega
\end{equation}
which is a fundamental principle in statistical mechanics.
Here, $k_{\rm B}$ represents the Boltzmann constant, approximately equal to $1.38\times 10^{-23} J/K$, and $\Omega$ denotes the number of microscopic states in the corresponding thermodynamic system. By rearranging this relationship, one obtains
\begin{equation}
 \Omega=e^{\frac{S}{k_{\rm B}}}\,,\label{OSs}
\end{equation}
which is the starting point of thermodynamic fluctuation theory. For a system described by $N+1$ independent variables $x^{\mu}$  ($\mu$=0, 1, ..., $N$), the probability of finding the system's state within a specific range defined by $(x^{0}, ..., x^{\rm N})$ and $( x^{0}+ dx^{0}, ..., x^{\rm N} + dx^{\rm N})$ is proportional to the number of microstates within that range. This probability can be expressed as
\begin{equation}
 P(x^{0}, ..., x^{\rm N})dx^{0}\cdot \cdot \cdot dx^{\rm N}=
 C\Omega (x^{0}, ..., x^{\rm N})dx^{0}\cdot \cdot \cdot dx^{\rm N}\,,
\end{equation}
where $C$ is a normalization constant. Hence
\begin{equation}
  P(x^{0}, ..., x^{\rm N})  \propto e^{\frac{S}{k_{\rm B}}}\,.
  \label{pp}
\end{equation}
Now, let us consider a thermodynamic system divided into a small subsystem $S$ and its environment $E$. The total entropy can be decomposed as $S(x^{0}, ..., x^{\rm N})=S_{\rm S}(x^{0}, ..., x^{\rm N})+S_{\rm E}(x^{0}, ..., x^{\rm N})$ with $S_{\rm S}\ll S_{\rm E}\sim S$. Expanding the total entropy near its local maximum at $x^{\mu} = x^{\mu}_0$, one obtains
\begin{align}
&S=S_{0} + \left. \frac{\partial S_{\rm S}}{\partial x^{\mu}}  \right|_{0} \Delta x^{\mu}_{\rm S}
+ \left.\frac{\partial S_{\rm E}}{\partial x^{\mu}}   \right|_{0}   \Delta x^{\mu}_{\rm E}
\label{qwq}\\
&+ \left. \frac{1}{2}\frac{\partial^{2}S_{\rm S}}{\partial x^{\mu}\partial x^{\nu}}
        \right|_{0}  \Delta x^{\mu}_{\rm S}\Delta x^{\nu}_{\rm S}
       + \left. \frac{1}{2}\frac{\partial^{2}S_{\rm E}}{\partial x^{\mu}\partial x^{\nu}}
        \right|_{0}  \Delta x^{\mu}_{\rm E}\Delta x^{\nu}_{\rm E}
   +\cdots \nonumber
\end{align}
where $S_{0}$ represents the local maximum of the entropy and ``$|_{0}$'' denotes evaluation at $|_{x^{\mu} = x^{\mu}_0}$.
It is reasonable to assume that the fluctuating parameters are both conserved and additive, meaning $x^{\mu}_{\rm S}+x^{\mu}_{\rm E}=x^{\mu}_{\rm total}$, which remains constant. Consequently, the sum of the second and third terms equals zero (or equivalently,
$\frac{\partial S_{\rm S}}{\partial x^{\mu}} |_{0} \Delta x^{\mu}_{\rm S}$ is the negative of $\frac{\partial S_{\rm E}}{\partial x^{\mu}} |_{0} \Delta x^{\mu}_{\rm E}$). Given that $S_{\rm E}$ is comparable in scale to the entire system, the final term in equation (\ref{qwq}) is substantially smaller than the fourth term and can therefore be neglected. Therefore, one can approximate the entropy change as
\begin{eqnarray}
 \Delta S=S-S_{0} = \left. \frac{1}{2}\frac{\partial^{2}S_{\rm S}}{\partial x^{\mu}\partial x^{\nu}}\right|_{0}  \Delta x^{\mu}_{\rm S}\Delta x^{\nu}_{\rm S} +\cdots\,.
\end{eqnarray}
Absorbing $S_{0}$ into the normalization constant, the probability expressed as
\begin{equation}
 P(x^{0}, ..., x^{\rm N}) \propto e^{-\frac{1}{2}\Delta l^{2}}\,,
\end{equation}
where $\Delta l^{2}$ represents the distance separating two adjacent fluctuation states, and is defined as \cite{Ruppeiner:1995zz,Ruppeiner:2012uc,Ruppeiner:2013yca}
\begin{eqnarray}\label{Ds}
 \Delta l^{2}=-\frac{1}{k_{\rm B}}\frac{\partial^{2}S_{\rm S}}{\partial x^{\mu}\partial x^{\nu}} \Delta x^{\mu}\Delta x^{\nu}\,.
\end{eqnarray}
In thermodynamic information geometry, the rarity of a fluctuation between two thermodynamic states is indicative of their greater separation. Consequently, the line element \eqref{Ds} contains information regarding the effective interaction that exists between two microscopic fluctuation states. For a specific fluid system, the curvature scalar derived from Eq.\eqref{Ds} serves as a gauge for its microstructure interactions. Notably, the entropy within this information geometry framework, known as Ruppeiner geometry, constitutes the thermodynamic potential.

To explore the microstructure of the black hole, one can utilize the principles of Ruppeiner geometry. For simplicity, we set the Boltzmann constant $k_{\rm B}$ to $1$ and omit the subscript $S$ from the entropy $S_{\text{S}}$. Applying the thermodynamic first law and considering temperature $T$ and thermodynamic volume $V$ as the fluctuation variables, the line element (\ref{Ds}) can be formulated as described in \cite{Wei:2019uqg}
\begin{equation}
 dl^{2}=\frac{C_{V}}{T^{2}}dT^{2}+\frac{(\partial_{V}P)_{T}}{T}dV^{2}\,,
 \label{dlTV}
\end{equation}
where
$C_{V}=T\big({\partial S}/{\partial T}\big)_{V}$ being the heat capacity at constant volume.

\section{$P$-$V$ Criticality and Microstructure of Charged AdS Black Hole with Quantum Anomaly for ${\alpha_{\rm c}} \leqslant 0$}\label{Rup-BHanomaly}

The thermodynamic geometry method offers a valuable instrument for exploring the microstructure of black holes. In this section, we conduct a comprehensive analysis of the thermodynamic properties and Ruppeiner geometry for a charged AdS (Anti-de Sitter) black hole exhibiting quantum conformal anomaly. Our approach begins by deriving the volumes of the coexisting small and large phases of this black hole, focusing on the region near the critical point. Subsequently, we delve into the behavior of the thermodynamic curvature for the black hole, specifically within the context of these coexisting volumes.

\subsection{Coexistence curves of the $P$-$V$ phase transition}

For the charged AdS black hole with quantum conformal anomaly, as described by Eq.(\ref{fr}), the thermodynamic equation of state is provided in \cite{Hu:2024ldp} and is given by
\begin{equation}\label{EoSPVT}
 P=\frac{T}{2 r_{\rm h}}-\frac{1}{8 \pi  r_{\rm h}^2}-\frac{2\alpha_{\rm c}T}{r_{\rm h}^3}+\frac{-2\alpha_{\rm c} +Q^2}{8\pi r_{\rm h}^4}\,,
\end{equation}
where $r_{\rm h}$ denotes the horizon radius and the thermodynamic volume $V$ is defined as $V=4\pi r_{\rm h}^{3}/3$.

\begin{figure}[h]
\centering
\includegraphics[scale=0.6]{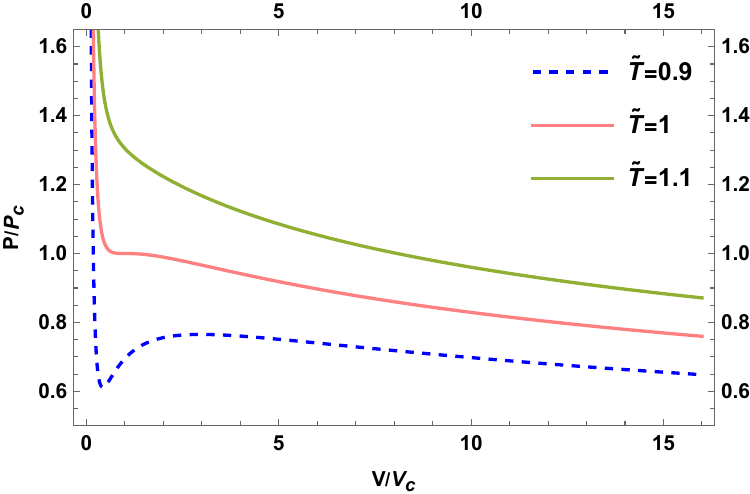}
\caption{Isotherms in the $\widetilde{P}$-$\widetilde{V}$ plane for the reduced equation of state of a charged AdS black hole with quantum conformal anomaly, with $\widetilde{T}=T/T_{\rm c}$.
The green, red, and dashed blue curves correspond to $\widetilde{T}=1.1$, $\widetilde{T}=1$, and $\widetilde{T}=0.9$,  respectively, with $\alpha_{\rm c}=-0.5$ and $Q=1$ fixed throughout all plots.}
\label{FigPV}
\end{figure}

The necessary condition for the occurrence of a $P$-$V$ phase transition in the system described by Eq.(\ref{EoSPVT}) is that the first and second derivatives of the pressure $P$ with respect to the volume $V$ at constant temperature $T$ must both be zero
\begin{equation}
\left(\frac{\partial P}{\partial V}\right)_{T}=\left(\frac{\partial^2 P}{\partial V^2}\right)_{T}=0\,.\label{PVTc}
\end{equation}
These conditions yield a critical point solution, denoted by $T=T_{\rm c},\ P=P_{\rm c}$ and $V=V_{\rm c}$.
It is known that this system for $\alpha_{\rm c} \leqslant 0$ exhibits a small-large black hole phase transition, with the critical point given by \cite{Wei:2020poh,Hu:2024ldp}
\begin{eqnarray}\label{crit}
V_{\rm c}&=&\frac{4\pi}{3}(-12\alpha_{\rm c}+3Q^2 + K)^{3/2}\,,\nonumber\\	
T_{\rm c}&=&\frac{3Q^2 - K}{48\pi \alpha_{\rm c} \sqrt{-12\alpha_{\rm c} +3Q^2+ K}}, \nonumber\\
P_{\rm c}&=&\frac{-18\alpha_{\rm c} +6Q^2 + K}{24\pi(-12\alpha_{\rm c}+3Q^2+ K)^2}\,,	
\end{eqnarray}
where $K\equiv\sqrt{192\alpha_{\rm c}^2+9Q^4-96\alpha_{\rm c} Q^2}$ and $V_{\rm c}=4\pi r_{\rm hc}^3/3$, with $r_{\rm hc}$ being the critical horizon radius.
To conveniently and explicitly display the diagrams, we first define the dimensionless reduced pressure, reduced volume, and reduced temperature as follows
\begin{equation}\label{reduce}
\widetilde{P}\equiv\frac{P}{P_{\rm c}}\,, \quad\quad \widetilde{V}\equiv\frac{V}{V_{\rm c}}\,, \quad\quad \widetilde{T}\equiv\frac{T}{T_{\rm c}}\,,
\end{equation}
and rewrite (\ref{EoSPVT}) as
\begin{eqnarray}\label{EoSPVT1}
\widetilde{P}&=& \frac{\widetilde{T} T_{\rm c} \big[\sqrt[3]{\pi}(6 V_{\rm c} \widetilde{V})^{2/3} -16 \pi \alpha_{\rm c} \big]}{6 P_{\rm c} V_{\rm c} \widetilde{V}} \\
&&+\frac{2 (6\pi)^{2/3} (Q^2-2 \alpha_{\rm c})-3 \sqrt[3]{6} (V_{\rm c}\widetilde{V})^{2/3}}{36 \sqrt[3]{\pi} P_{\rm c} (V_{\rm c} \widetilde{V})^{4/3}}\,.\nonumber
\end{eqnarray}
The corresponding critical behavior of the black hole is illustrated in Fig.\ref{FigPV}.
The diagram reveals that for
$T>T_{\rm c}$, the curves exhibit ideal gas behavior, indicating a unique phase of the black hole. For $T<T_{\rm c}$, the oscillatory behavior suggests the existence of small and large black hole phases, which undergo a first-order phase transition. The red curve (at $T=T_{\rm c}$) displays the expected inflection point characteristic of a critical point.

To observe the divergence behavior of the thermodynamic curvature of charged AdS black hole with quantum conformal anomaly in the next subsection, we provide the reduced coexisting volumes during the $P$-$V$ phase transition.
For this goal, we make an expansion to equation of state (\ref{EoSPVT}) near the critical point as
\begin{equation}\label{seriesP}
\widetilde{P}(t,\omega)=1+a_{10}t+a_{11}t\omega+a_{03}\omega^3+\mathcal{O}(t\omega^3,\omega^4)
\end{equation}
where $t\equiv \widetilde{T}-1$, $\omega\equiv\widetilde{V}-1$, and the coefficients are given by \cite{Hu:2024ldp}
\begin{eqnarray}
 a_{10}&=&\frac{4(3Q^2-4\alpha_{\rm c}+K)}{9Q^2-22\alpha_{\rm c}}\,,\\
 a_{11}&=&\frac{2(3Q^2-5K-24\alpha_{\rm c})}{27Q^2-66\alpha_{\rm c}}\,,\,\,\nonumber
 a_{03}=\frac{K-15Q^2+40\alpha_{\rm c}}{27(9Q^2-22\alpha_{\rm c})}\,.\,\,\nonumber
\end{eqnarray}
When the black hole system undergoes a phase transition from a small black hole to a large one, the temperature and pressure remain constant, while the thermodynamic volume changes from $\omega_s$ to $\omega_l$, where the labels `s' and `l' represent `small' and `large' respectively. During the phase transition, Maxwell's equal area law holds \cite{Spallucci:2013osa,Lan:2015bia,Xu:2015hba} (see also Refs. \cite{Majhi:2016txt,Bhattacharya:2017hfj,Abdusattar:2023pck} for more related discussion)
\begin{equation}\label{m1}
       \widetilde{P}(t,\omega_{s})=\widetilde{P}(t,\omega_{l})\,,~~~~
       \int_{\omega_{s}}^{\omega_{l}} \omega d\widetilde{P}=0\,.
   \end{equation}

\begin{figure}[h]
\centering
\includegraphics[scale=0.6]{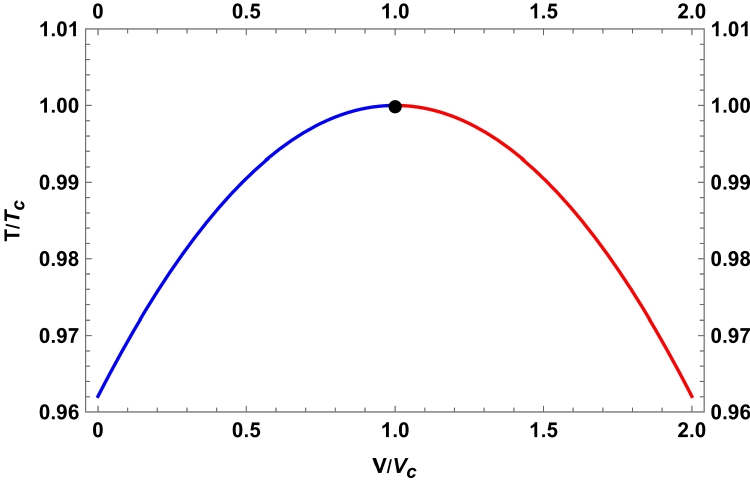}
\caption{Coexistence curves of small and large black hole phases in the $\widetilde{T}$-$\widetilde{V}$ diagram for charged AdS black holes with quantum anomaly. Here we set $\alpha_{\rm c}=-0.5, Q=1$.}
\label{FigTV}
\end{figure}

Employing Maxwell's equal area law, we obtain the non-trivial reduced volumes of the coexistence small and large phases of the charged AdS black hole with quantum conformal anomaly near the critical point, which are
\begin{eqnarray}
\widetilde{V}_s&=&1-\sqrt{\frac{18K}{3Q^2-8\alpha_{\rm c}}(1-\widetilde{T})}\,,\,\,\,\,\,\label{Vs} \\
\widetilde{V}_l&=&1+\sqrt{\frac{18K}{3Q^2-8\alpha_{\rm c}}(1-\widetilde{T})}\,.\,\,\label{Vl}
\end{eqnarray}
The coexistence curves in the $\widetilde{T}$-$\widetilde{V}$ diagram corresponding to Eqs.(\ref{Vs}) and (\ref{Vl}) are shown in Fig.\ref{FigTV}.

The coexistence curves provide a clear visual representation of the phase transition behavior near the critical point. Specifically, the blue (left) solid curve represents small black holes, while the red (right) solid curve represents large black holes. These curves meet at the critical point, denoted by a black dot, and the area beneath these curves signifies the coexistence phase.


\subsection{Thermodynamic curvature via $P$-$V$ criticality}

In the following, we apply the Ruppeiner geometry to investigate the divergence behavior of the thermodynamic curvature of charged AdS black hole with quantum anomaly along the coexisting volumes during the $P$-$V$ phase transition.

One special property of charged AdS black hole with quantum anomaly is that their heat capacity at constant volume vanishes, $i.e.$, $C_{V}$=0. This leads the line element (\ref{dlTV}) singular, and consequently information of the associated black hole microstructure is not revealed from the thermodynamic geometry. To avoid this problem, one can treat $C_{V}$ as a constant whose value is infinitesimally close to zero, defining a new normalized scalar curvature \cite{Wei:2019uqg,Wei:2019yvs}
\begin{eqnarray}\label{RNNSC}
R_{\rm N}&=&C_{V} R\\
&=& \frac{(\partial_V P)^2 - T^2(\partial_{V, T} P)^2 + 2T^2(\partial_V P)(\partial_{V, T, T} P)}{2(\partial_V P)^2}\,.\nonumber
\end{eqnarray}
%
By utilizing Equations (\ref{EoSPVT}), (\ref{crit}), and (\ref{reduce}), we derive the normalized scalar curvature for the charged AdS black hole incorporating quantum anomaly, as outlined below
\begin{eqnarray} \label{RN1}
 R_{\rm N}=\frac{2 \delta \beta_0}{(\beta_0 +\delta)^2}\,,
%
\end{eqnarray}
where
\begin{eqnarray}
 \beta_0 &=&\delta +\widetilde{T} \widetilde{V} (K^2-9 Q^4)-12 \alpha_{\rm c} (K-3 Q^2)(1+\widetilde{V}^{2/3}) \widetilde{T}{\widetilde{V}}^{1/3} \,\nonumber  \\
 \delta &=&12 \alpha_{\rm c} [4 \alpha_{\rm c} +\widetilde{V}^{2/3} (K-12 \alpha_{\rm c})+Q^2 (3 \widetilde{V}^{2/3}-2)]\,.\nonumber
\end{eqnarray}
Note that $R_{\rm N}$ explicitly depend on the black hole charge $Q$ and quantum anomaly. As $\widetilde{T}$ tends to $0$, we observe that $R_{\rm N}$ converges to $1/2$ for both phases, reflecting a common asymptotic property in the low-temperature limit.
However, in proximity to the critical point defined by the equation
\begin{eqnarray}\label{spcurve}
\widetilde{T}_{\rm div}=\frac{24 \alpha_{\rm c} [4 \alpha_{\rm c} +\widetilde{V}^{2/3} (K-12 \alpha_{\rm c})+Q^2 (3 \widetilde{V}^{2/3}-2)]}{\widetilde{V} (9 Q^4-K^2)+12 \alpha_{\rm c} (\widetilde{V}^{1/3}+\widetilde{V}) (K-3 Q^2)}\,\,\,\,\,\,\,\,
\end{eqnarray}
$R_{\rm N}$ undergoes significant variations, diving to negative infinity along this curve. This behavior suggests a rapid transformation in the microstructure of the charged AdS black hole with quantum anomaly near the temperature $\widetilde{T}_{\rm div}$. Furthermore, we have computed the curves where $R_{\rm N}$ changes its sign, given by
\begin{eqnarray}\label{sc1}
 \widetilde{T}_{0}=\frac{\widetilde{T}_{\rm div}}{2}\,,
\end{eqnarray}
whose traversal indicates a change between attractive or repulsive interactions within the microstructure.

Finally, we discuss the behavior of the thermodynamic curvature as it approaches the critical point along the coexistence curve. To this end, we first express the normalized scalar curvature $R_{\rm N}$ as a function of the reduced temperature $\widetilde{T}$ by substituting Eqs. (\ref{Vs}), (\ref{Vl}), and (\ref{RN1}); we then derive its series expansions along the coexistence curve, which take the following forms
\begin{eqnarray}
R_{\rm N}^{s}&=&-\frac{1}{8 t^2}-
\frac{\eta_0}{{[(8\alpha_{\rm c} -3 Q^2) t]}^{3/2} }+...\mathcal{O}(t^{-1}, ...)\,, \label{RNs}\\
R_{\rm N}^{l}&=&-\frac{1}{8 t^2}+
\frac{\eta_0}{{[(8\alpha_{\rm c}-3 Q^2) t]}^{3/2} }+...\mathcal{O}(t^{-1}, ...)\,, \label{RNl}
\end{eqnarray}
with
\begin{equation}
    \eta_0=\frac{\sqrt{K} (768\alpha_{\rm c}^2-80\alpha_{\rm c} K+31 K Q^2+63 Q^4-456\alpha_{\rm c} Q^2)}{4\sqrt{2}(-8 \alpha_{\rm c} +K+Q^2)}
\end{equation}
where $t=\widetilde{T}-1$ is the deviation from the critical temperature. We see that $R_{\rm N} \rightarrow - \infty$ at the critical point with a universal critical exponent of $2$.

\begin{figure}[h]
\includegraphics[scale=0.7]{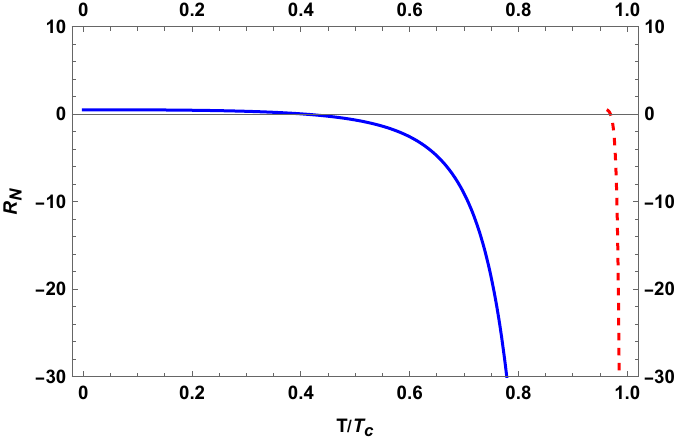}
\caption{The normalized thermodynamic curvature $R_{\rm N}$ of the charged AdS black hole with quantum anomaly for $\alpha_{\rm c} < 0$ along the coexistence small and large phases. Here we set $\alpha_{\rm c}=-0.5, Q=1$.}
\label{FigRNT1}
\end{figure}

Furthermore, by utilizing Eqs.(\ref{RNs}) and (\ref{RNl}) and ignoring the high orders, we obtain the following expression
\begin{equation}\label{universal C}
 \lim_{t\rightarrow 0} R_{\rm N} t^2=-\frac{1}{8}\,.
\end{equation}
This analysis unveils a universal, dimensionless constant of $-1/8$, which aligns perfectly with the numerical findings for the van der Waals fluid, as reported in \cite{Wei:2019uqg}. To visualize how $R_{\rm N}$ diverges, we present a graphical representation of the normalized thermodynamic curvature for both the small and large coexistence phases in Fig.\ref{FigRNT1}.

Figure \ref{FigRNT1} shows that the normalized thermodynamic curvature $R_{\rm N}$ in both small and large phases diverges to negative infinity ($R_{\rm N} \rightarrow -\infty$) at the critical point. Specifically, for the small black hole phase (dashed red curve), $R_{\rm N}$ changes sign to positive as the temperature deviates from the critical temperature.
For the large black hole phase (solid blue curve), $R_{\rm N}$ exhibits a distinct behavior, remaining nearly constant at small $\widetilde{T}$ when the temperature deviates from the critical temperature. This contrasts with a van der Waals fluid system, where weak repulsive interactions prevail throughout the parameter space above the coexistence curve.

\section{$P$-$V$ Criticality and Microstructure of Charged AdS Black Hole with Quantum Anomaly for $0 < \alpha_{c} < \frac{Q^{2}}{8}$}\label{sec:PV20}

We restricted our analysis in the previous section to the case where $\alpha_{c} \leqslant 0$. In contrast, accounting for the non-negativity of the central charge and the existence of a critical point narrows the physically viable parameter region to $0 < \alpha_{c} \leqslant \frac{Q^{2}}{8}$.
Therefore, we focus our subsequent investigations on the phase transitions and microstructure of the black hole in this regime.


\subsection{Coexistence curves of the $P$-$V$ phase transition}

\begin{figure}[h]
\centering
\includegraphics[scale=0.7]{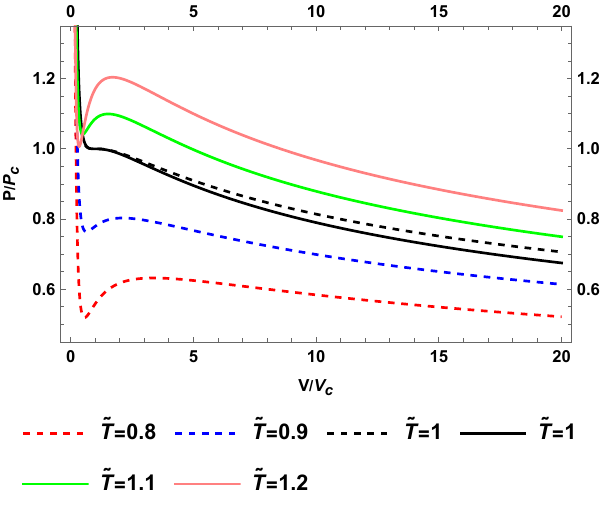}
\caption{$\widetilde{P}$-$\widetilde{V}$ isotherms for the reduced equation of state of a quantum-anomalous charged AdS black hole. Curves correspond to normalized temperatures $\widetilde{T}=T/T_{\rm c}$:
dashed red ($\widetilde{T}=0.8$), dashed blue ($\widetilde{T}=0.9$), dashed black ($\widetilde{T}=1$), solid black ($\widetilde{T}=1$), solid green ($\widetilde{T}=1.1$), solid red ($\widetilde{T}=1.2$). Dashed curves are plotted via Eq.(\ref{crit}), while solid curves are via Eq.(\ref{crit20}). Parameters are set as $\alpha_{\rm c}=0.1$ and $Q=1$.}
\label{FigPV20}
\end{figure}

In this physical case, in addition to the critical point obtained in (\ref{crit}), there is another critical point \cite{Hu:2024ldp}
\begin{eqnarray}\label{crit20}
V_{\rm c}&=&\frac{4\pi}{3}(-12\alpha_{\rm c}+3Q^2 - K)^{3/2}\,,\nonumber\\	
T_{\rm c}&=&\frac{3Q^2 + K}{48\pi \alpha_{\rm c} \sqrt{-12\alpha_{\rm c} +3Q^2- K}}\,, \nonumber\\
P_{\rm c}&=&\frac{-18\alpha_{\rm c} +6Q^2 - K}{24\pi(-12\alpha_{\rm c}+3Q^2- K)^2}\,.
\end{eqnarray}
It is worth noting that when $\alpha_{\rm c}$ tends to zero, the critical point (\ref{crit20}) shows a divergent behavior and is therefore not considered herein, while the critical values presented in Eq. (\ref{crit}) reduce to those of the Reissner-Nordstr\"{o}m-Anti-de Sitter (RN-AdS) black hole \cite{Kubiznak:2012wp}.
However, when the central charge $\alpha_{\rm c}$ and the electric charge parameter $Q$ satisfy the specific relation $\alpha_{\rm c}=Q^2 /8$, the auxiliary quantity $K$ (appearing in both Eqs. (\ref{crit}) and (\ref{crit20})) becomes zero, and the two critical points merge into one--this intriguing specific case will be investigated in detail in the next section.
Therefore, we focus our discussion here on the critical behavior of quantum-anomalous charged AdS black holes with a central charge in the range $0<\alpha_{\rm c}<Q^2 /8$, which is demonstrated based on Eqs.(\ref{EoSPVT1}), (\ref{crit}) and (\ref{crit20}) and shown in Fig. \ref{FigPV20}.

In Fig.\ref{FigPV20}, both the dashed and solid curves are derived from the equation of state (\ref{EoSPVT1}): the former uses the critical point in Eq.(\ref{crit}), and the latter uses the alternative critical point in Eq.(\ref{crit20}). Notably, phase transitions associated with the dashed curves occur only for temperatures below the critical temperature, whereas those corresponding to the solid curves emerge for temperatures above the critical temperature.\footnote{This behavior differs from that of van der Waals systems and most black hole systems, where coexisting phases typically appear below the critical temperature \cite{Kubiznak:2012wp,Altamirano:2014tva,Kubiznak:2016qmn}.} Additionally, the system exhibits a ``thermodynamic singularity"\footnote{This singularity manifests as a common intersection point for different isothermal lines in the $\widetilde{P}$-$\widetilde{V}$ diagram, a feature that has also been identified in some previous studies on black hole thermodynamics \cite{Frassino:2014pha,Hennigar:2016ekz}.}, defined by the temperature independence of pressure, $i.e.$, $(\partial \widetilde{P}/\partial \widetilde{T})_{\widetilde{V}}=0$, see more detailed discussion in Appendix \ref{Appendix}.

\begin{figure}[h]
\centering
\includegraphics[scale=0.7]{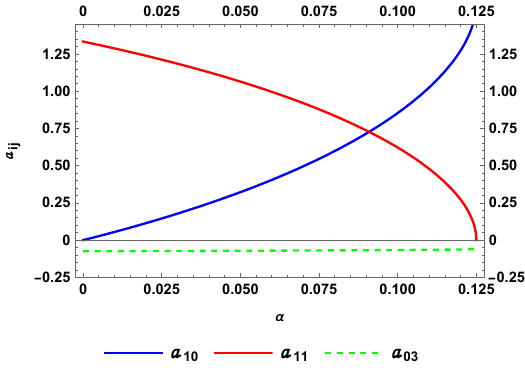}
\caption{The coefficients $a_{ij}$ as a function of $\alpha_{\rm c}$ in the range $(0, {Q^{2}}/{8}]$. Here, we set $Q=1$.}\label{Figa}
\end{figure}

By expanding the equation of state (\ref{EoSPVT}) around the critical point (\ref{crit20}) via a Taylor series, we obtain the following result
\begin{equation}\label{seriesP20}
\widetilde{P}(t,\omega)=1+a_{10}t+a_{11}t\omega+a_{03}\omega^3+...\mathcal{O}(t\omega^3,\omega^4)
\end{equation}
where the coefficients $a_{10}$, $a_{11}$, and $a_{03}$ in Eq.(\ref{seriesP20}) are given by \cite{Hu:2024ldp} as follows
\begin{eqnarray}
a_{10}&=&\frac{4(3Q^2-4\alpha_{\rm c}-K)}{9Q^2-22\alpha_{\rm c}}\,,\\
a_{11}&=&\frac{2(5K+3Q^2-24\alpha_{\rm c})}{27Q^2-66\alpha_{\rm c}}\,,\,\,\nonumber
a_{03}=\frac{40\alpha_{\rm c}-K-15Q^2}{27(9Q^2-22\alpha_{\rm c})}\,.\,\,\nonumber
\end{eqnarray}
The variation of these coefficients with $\alpha_{\rm c}$ is visualized in Fig.\ref{Figa}.

\begin{figure}[h]
\centering
\includegraphics[scale=0.7]{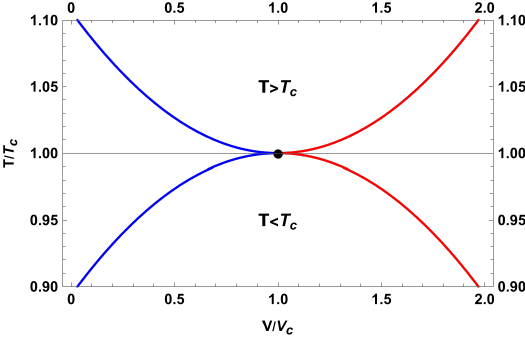}
\caption{Coexistence curves:
$\widetilde{T}$-$\widetilde{V}$ phase structure of the quantum-anomalous charged AdS black hole. Here, we set $\alpha_{\rm c}=0.1$ and $Q=1$.}
\label{FigTV20}
\end{figure}

Following the same approach as in the previous section, we employ Maxwell's equal-area law and, from Eq.(\ref{seriesP20}), derive the non-trivial reduced volumes corresponding to the coexisting small and large black hole phases of the quantum-anomalous charged AdS black hole near the critical point defined in Eq.(\ref{crit20}).
These volumes are given by\footnote{The difference between the reduced volumes of the large and small black hole phases
$\widetilde{V}_{l}-\widetilde{V}_{s}\propto |t|^{{1}/{2}}$,
which gives one of the critical exponents ($1/2$) for the coexistence curve in the context of scaling laws.}
\begin{eqnarray}
\widetilde{V}_s&=&1-\sqrt{\frac{18K}{3Q^2-8\alpha_{\rm c}}(\widetilde{T}-1)}\,,\,\,\,\,\,\label{Vs20} \\
\widetilde{V}_l&=&1+\sqrt{\frac{18K}{3Q^2-8\alpha_{\rm c}}(\widetilde{T}-1)}\,.\,\,\label{Vl20}
\end{eqnarray}
To gain a deeper understanding of the phase transition behavior of these coexisting small and large phases in the vicinity of the critical point, we plot the temperature-volume diagram ($\widetilde{T}$-$\widetilde{V}$) based on Eqs. (\ref{Vs}), (\ref{Vl}), (\ref{Vs20}) and (\ref{Vl20}); this diagram is presented in Fig. \ref{FigTV20}.

As observed in Fig. \ref{FigTV20}, the small and large black hole phases are located on the left and right sides of the $\widetilde{T}$-$\widetilde{V}$ diagram, respectively. The blue and red curves on either side correspond to these two phases, with both curves converging at the critical point (denoted by a black dot). Notably, the coexistence of these two phases is evident for both
$T<T_{\rm c}$ and $T>T_{\rm c}$, which indicates that the small and large black hole phases undergo first-order phase transitions both below and above the critical temperature.


\subsection{Thermodynamic curvature via $P$-$V$ criticality}

By substituting Eqs. (\ref{EoSPVT}), (\ref{reduce}), and (\ref{crit20}) into Eq. (\ref{RNNSC}), we derive the normalized scalar curvature $R_{\rm N}$ for the quantum-anomalous charged AdS black hole within the parameter range $0< \alpha_{\rm c} <{Q^{2}}/{8}$. One of the resulting expressions is given as
\begin{eqnarray} \label{RN20}
R_{\rm N}=\frac{24 \alpha_{\rm c} (4\alpha_{\rm c} -2 Q^2+\eta {\widetilde{V}}^{2/3}) [\Theta +12 \alpha_{\rm c} (4\alpha_{\rm c} +\eta {\widetilde{V}}^{2/3})]}{[\Theta +24 \alpha_{\rm c} (4 \alpha_{\rm c}-Q^2+\eta  {\widetilde{V}}^{2/3})]^2} \,\,\,\,\,
\end{eqnarray}
where the auxiliary quantities $\Theta$ and $\eta$ are defined as
\begin{eqnarray}
\Theta&=& K \widetilde{T} (12 \alpha_{\rm c} {\widetilde{V}}^{1/3}-\eta \widetilde{V})-3 Q^2 [(8-12 \widetilde{T} {\widetilde{V}}^{1/3})\alpha_{\rm c} +\eta \widetilde{T} \widetilde{V}] \,,\nonumber\\
\eta&=&3 Q^2-12\alpha_{\rm c} -K \,. \nonumber
\end{eqnarray}
Notably, $R_{\rm N}$ exhibits an explicit dependence on the black hole's electric charge $Q$ and central charge $\alpha_{\rm c}$. In the low-temperature limit ($\widetilde{T}\rightarrow 0$), this quantity also converges to a constant value of $1/2$.
In the vicinity of the curve defined by
\begin{eqnarray}\label{spcurve20}
\widetilde{T}_{\rm div}=\frac{24 \alpha_{\rm c} [Q^2 (2-3 \widetilde{V}^{2/3})-4 \alpha_{\rm c} +\widetilde{V}^{2/3} (K+12 \alpha_{\rm c})]}{\widetilde{V} (K^2-9 Q^4)+12 \alpha_{\rm c} (\widetilde{V}^{1/3}+\widetilde{V}) (K+3 Q^2)}\,\,\,\,
\end{eqnarray}
$R_{\rm N}$ undergoes a dramatic variation, tending to negative infinity exactly on this curve. This abrupt change implies that the microstructure of the black hole undergoes significant alterations near the temperature $\widetilde{T}_{\rm div}$. Furthermore, we calculate the curves where $R_{\rm N}$ changes sign, which are given by
\begin{eqnarray}\label{sc20}
 \widetilde{T}_{0}=\frac{\widetilde{T}_{\rm div}}{2}\,,
\end{eqnarray}
indicating that traversing these curves corresponds to a transition between attractive and repulsive interactions among the black hole's microstates.

Finally, we discuss the behavior of the thermodynamic curvature as it approaches the critical point along the coexistence curve. To this end, we express the normalized scalar curvature
$R_{\rm N}$ as a function of $\widetilde{T}$ by substituting Eqs. (\ref{Vs20}), (\ref{Vl20}), and (\ref{RN20}); we then derive the series expansions of $R_{\rm N}$ at $\widetilde{T}$ along the coexistence curve for the small and large black hole phases, respectively, which are given by
\begin{eqnarray}
R_{\rm N}^{s}&=&-\frac{1}{8 t^2}-
\frac{[3( Q^2-8\alpha_{\rm c})]\xi}{{[(3 Q^2-8\alpha_{\rm c})]}^{5/4} t^{3/2}}+...\mathcal{O}(t^{-1})\,,\,\,\, \label{RNs20}\\
R_{\rm N}^{l}&=&-\frac{1}{8 t^2}+
\frac{[3( Q^2-8\alpha_{\rm c})]\xi}{{[(3 Q^2-8\alpha_{\rm c})]}^{5/4} t^{3/2}}+...\mathcal{O}(t^{-1})\,,\,\,\, \label{RNl20}
\end{eqnarray}
where $\xi$ is defined as
\begin{equation}
\xi=\frac{768\alpha_{\rm c}^2+80\alpha_{\rm c} K-31 K Q^2+63 Q^4-456\alpha_{\rm c} Q^2}{4\sqrt{2}(8 \alpha_{\rm c} +K-Q^2)}\,.\nonumber
\end{equation}
From these expansions, we observe that
$R_{\rm N}$ tends to $-\infty$ at the critical point, with a universal critical exponent of $2$. Furthermore, by substituting Eqs. (\ref{RNs20}) and (\ref{RNl20}) and neglecting higher-order terms, we obtain the following limiting expression
\begin{equation}\label{universalC20}
 \lim_{t\rightarrow 0} R_{\rm N} t^2=-\frac{1}{8}\,.
\end{equation}
This analysis reveals a universal dimensionless constant of $-1/8$, which is in perfect agreement with the numerical results for the van der Waals fluid reported in \cite{Wei:2019uqg}.

\begin{figure}[h]
\includegraphics[scale=0.7]{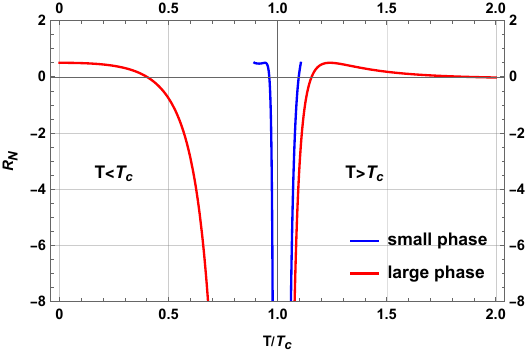}
\caption{Normalized thermodynamic curvature $R_{\rm N}$ as a function of $\widetilde{T}=T/T_{\rm c}$ for quantum-anomalous charged AdS black holes along their coexistence phases. The blue and red curves correspond to the small and large black hole phases, respectively; both phases exist for $T<T_{\rm c}$ and $T>T_{\rm c}$, with $R_{\rm N}$ diverging to $-\infty$ as $\widetilde{T}\rightarrow 1$ ($i.e., T\rightarrow T_{\rm c}$). Here we set $\alpha_{\rm c}=0.1, Q=1$.}
\label{FigRNT20}
\end{figure}

To visualize the divergent behavior of the normalized thermodynamic curvature $R_{\rm N}$ for quantum-anomalous charged AdS black holes (within the parameter range $0<\alpha_{c}<{Q^{2}}/{8}$), we plot $R_{\rm N}$ along the respective coexistence curves of the small and large black hole phases in Fig.\ref{FigRNT20}, based on Eqs.(\ref{RN1}) and (\ref{RN20}).

Figure \ref{FigRNT20} demonstrates that at the critical point, the normalized thermodynamic curvature $R_{\rm N}$ diverges to negative infinity ($R_{\rm N} \rightarrow -\infty$) in both the small and large phases.
It is noteworthy that for temperatures below ($T<T_{\rm c}$) and above ($T>T_{\rm c}$) the critical temperature, $R_{\rm N}$ can be positive as the temperature deviates from the critical temperature, implying that the microstructural interaction in the coexistence small and large phases of charged AdS black holes with quantum anomalies undergoes a transition from attractive to repulsive during the phase transition.

\section{$P$-$V$ Criticality and Microstructure of Charged AdS Black Hole with Quantum Anomaly for $\alpha_{\rm c}={Q^{2}}/{8}$}\label{RupBHaQ}

In the context of four-dimensional black hole system, it has been observed that under specific conditions related to the central charge and $U(1)$ charge ($\alpha_{\rm c}={Q^{2}}/{8}$), the coefficient $a_{11}$ in an expansion of the equation of state near the critical point can vanish, $i.e.$, $a_{11}=0$, results in critical exponents that deviate from those predicted by mean field theory, instead adhering to unique scaling laws outlined in \cite{Hu:2024ldp}. This distinctive characteristic sets the black hole exhibiting quantum anomalies apart from a van der Waals fluid. Consequently, it is intriguing to explore whether the thermodynamic scalar curvature of such a black hole system differs from that of a van der Waals system.



\subsection{Coexistence curves of the $P$-$V$ phase transition}

\begin{figure}[h]
\centering
\includegraphics[scale=0.6]{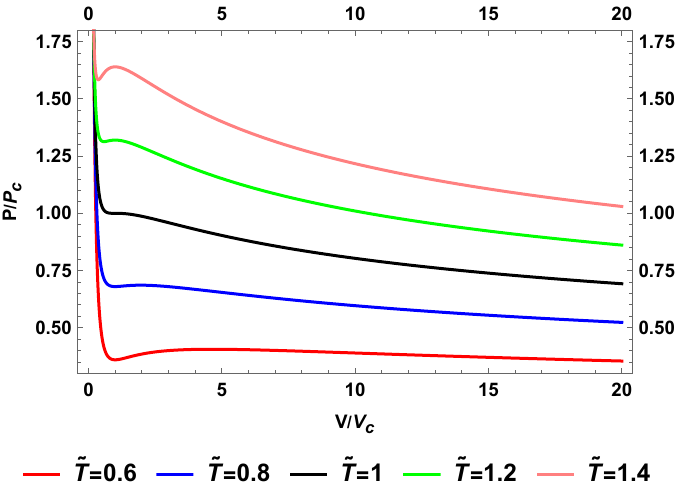}
\caption{$\widetilde{P}$-$\widetilde{V}$ isotherms for the reduced equation of state of charged AdS black hole with quantum anomaly \cite{Hu:2024ldp}. It shows the oscillatory behaviors for both of $T<T_{\rm c}$ and $T>T_{\rm c}$, indicating the existence of small and large black hole phases which undergo a first-order phase transitions occur both above and below the critical temperature. Here we set $Q=1$.}
\label{FigPV2}
\end{figure}

The equation of state with the specific condition involving the central charge and $U(1)$ charge ($\alpha_{\rm c}={Q^{2}}/{8}$) reads \cite{Hu:2024ldp}
\begin{equation}\label{EoSPVT2}
 P=\frac{T}{2 r_{\rm h}}-\frac{1}{8 \pi r_{\rm h}^2}+\frac{Q^2 (8 \pi r_{\rm h} T+3)}{32 \pi r_{\rm h}^4}\,,
\end{equation}
where the thermodynamic volume $V=4\pi r_{\rm h}^{3}/3$.
From the criticality condition (\ref{PVTc}), one obtains the critical point as
\begin{eqnarray}\label{crit2}
V_{\rm c}=\sqrt{6}\pi Q^{3}\,, \quad T_{\rm c}=\frac{1}{\sqrt{6}\pi Q}\,, \quad P_{\rm c}=\frac{5}{72\pi Q^{2}}\,.	
\end{eqnarray}
For this special case, the critical behavior of charged AdS black hole with quantum anomaly is shown in Fig.\ref{FigPV2}.

\begin{figure}[h]
\centering
\includegraphics[scale=0.6]{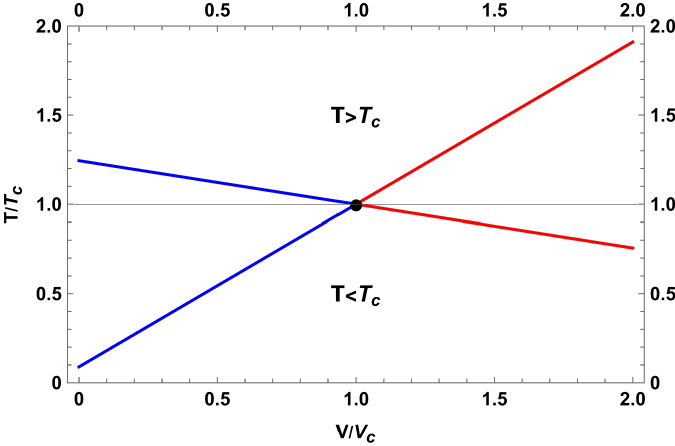}
\caption{Coexistence curves: phase structure of the charged AdS black hole with quantum anomaly in $\widetilde{T}$-$\widetilde{V}$ diagram.}
\label{FigTV2}
\end{figure}

Expand the equation of state (\ref{EoSPVT2}) near the critical point like Eq.(\ref{seriesP}), one can obtain \cite{Hu:2024ldp}
\begin{equation}\label{seriesP1}
\widetilde{P}(t,\omega)=1+\frac{8}{5}t-\frac{4}{15}t\omega-\frac{8}{135}\omega^3+...\mathcal{O}(t\omega^3,\omega^4) \,.
\end{equation}
Employing Maxwell's equal area law as the same method discussed above sections, we obtain the non-trivial reduced volumes of the coexistence small and large phases of black hole with quantum anomaly near the critical point, which are
\begin{eqnarray}
\widetilde{V}_s&=&1-\frac{3}{2}(\sqrt{3}+1)(\widetilde{T}-1)\,,\,\,\,\,\,\label{Vs2} \\
\widetilde{V}_l&=&1+\frac{3}{2}(\sqrt{3}-1)(\widetilde{T}-1)\,.\,\,\label{Vl2}
\end{eqnarray}
To gain a better understanding of the phase transition behaviors for the coexistence small and large phases around the critical point, we plot the $\widetilde{T}$-$\widetilde{V}$ diagram corresponding to Eqs.(\ref{Vs2}) and (\ref{Vl2}), as shown in Fig.\ref{FigTV2}.

We observe that the small and large black hole phases are positioned on the left and right, respectively, with the blue and red curves on either side depicting them and converging at the critical point (marked as a black dot). This coexistence phase is evident for both $T<T_{\rm c}$ and $T>T_{\rm c}$, indicating the presence of small and large black hole phases that undergo first-order phase transitions both above and below the critical temperature.

\subsection{Thermodynamic curvature via $P$-$V$ criticality}

Inserting Eqs.(\ref{reduce}), (\ref{EoSPVT2}) and (\ref{crit2}) into (\ref{RNNSC}), we derive the normalized scalar curvature for the charged AdS black hole incorporating quantum anomaly with $\alpha_{\rm c}={Q^{2}}/{8}$, as obtained below
\begin{eqnarray} \label{RN2}
 R_{\rm N}=\frac{1-2 \widetilde{T} {\widetilde{V}}^{1/3}}{2 (\widetilde{T} {\widetilde{V}}^{1/3}-1)^2} \,.
\end{eqnarray}
Note that $R_{\rm N}$ does not explicitly depend on the black hole charge $Q$--all black holes with different charge share the same expression in the reduced parameter space, a universal result. Moreover, as $\widetilde{T}\rightarrow 0$, our analysis reveals that $R_{\rm N}$ approaches the asymptotic value $1/2$.
In the proximity of the curve formulated by
\begin{eqnarray}\label{spcurve2}
 \widetilde{T}_{\rm div}=\frac{1}{{\widetilde{V}}^{1/3}}\,
\end{eqnarray}
$R_{\rm N}$ undergoes dramatic changes, approaching negative infinity on this curve. This abrupt change suggests that the microstructure of the black hole undergoes rapid alterations near the temperature $\widetilde{T}_{\rm div}$.
Furthermore, we calculate the curves where $R_{\rm N}$ changes its sign, given by
\begin{eqnarray}\label{sc2}
\widetilde{T}_{0}=\frac{\widetilde{T}_{\rm div}}{2}\,,
\end{eqnarray}
indicating a transition between attractive and repulsive interactions within the microstructure as one traverses these curves. Notably, these curves exhibit universality, as they apply to all charged AdS black holes (with quantum anomalies) regardless of the specific value of the electric charge $Q$.

Finally, we discuss the behaviors of thermodynamic curvature towards the critical point along the coexistence curve. For this purpose, we write $R_{\rm N}$ as a function of $\widetilde{T}$ by using Eqs.(\ref{Vs2}), (\ref{Vl2}) and (\ref{RN2}), the series expansions of it at $\widetilde{T}$ along the coexistence curves
\begin{eqnarray}
R_{\rm N}^{s}&=&\frac{-2-\sqrt{3}}{t^2}+\frac{34+20 \sqrt{3}}{t}+...\mathcal{O}(t^{1}, ...)\,, \label{RNs2}\\
R_{\rm N}^{l}&=&\frac{-2+\sqrt{3}}{t^2}+\frac{34-20 \sqrt{3}}{t}+...\mathcal{O}(t^{1}, ...)\,, \label{RNl2}
\end{eqnarray}
for the small and large black hole cases
respectively, where $t=\widetilde{T}-1$. We see that $R_{\rm N}\rightarrow - \infty$ at
the critical point with a universal critical exponent of $2$. Furthermore, by utilizing Eqs.(\ref{RNs2}) and (\ref{RNl2}) and ignoring the high orders, we obtain the following expression
\begin{equation}\label{universalC}
 \lim_{t\rightarrow 0} R_{\rm N} t^2=-2\pm\sqrt{3}\,.
\end{equation}
This result differs from the corresponding value obtained for a van der Waals fluid, which is$-1/8$ \cite{Wei:2019uqg,Wei:2019yvs}.

\begin{figure}[h]
\includegraphics[scale=0.7]{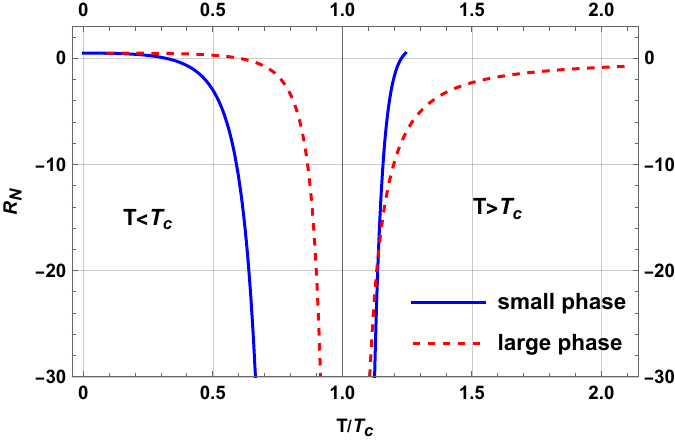}
\caption{Normalized thermodynamic curvature
$R_{\rm N}$ for charged AdS black holes with quantum anomaly along their coexistence phases. The blue curve represents the small black hole phase, and the dashed red curve represents the large black hole phase; both phases exist for $\widetilde{T}=T/T_{\rm c}<1 (i.e., T<T_{\rm c})$ and $\widetilde{T}=T/T_{\rm c}>1 (i.e., T>T_{\rm c})$.}
\label{FigRNT2}
\end{figure}

To see the divergence behavior of $R_{\rm N}$, we illustrate the normalized thermodynamic curvature along the coexistence small and large phases in Fig.\ref{FigRNT2}.

Figure \ref{FigRNT2} demonstrates that at the critical point, the normalized thermodynamic curvature $R_{\rm N}$ diverges to negative infinity ($R_{\rm N} \rightarrow -\infty$) in both the small and large phases. It is noteworthy that for temperatures below the critical temperature ($T<T_{\rm c}$), $R_{\rm N}$ changes sign to positive in both the small and large black hole phases, suggesting that a weak repulsive interaction dominates in charged AdS black holes with quantum anomalies at low temperatures.
For temperatures above the critical temperature ($T>T_{\rm c}$), $R_{\rm N}$ remains negative along the coexistence curve for the large black hole phase, indicating that an attractive interaction dominates in charged AdS black holes with quantum anomalies. However, along the coexistence curve for the small black hole phase, $R_{\rm N}$ can be positive as the temperature deviates from the critical temperature, implying that the microstructural interaction in the coexistence small phase of charged AdS black holes with quantum anomalies undergoes a transition from attractive to repulsive during the phase transition.


\section{Conclusions and Discussion}\label{conclusion}

In this paper, we adopt Ruppeiner thermodynamic geometry as a robust framework to investigate the microscopic structure of charged AdS black holes with quantum anomalies. Incorporating quantum effects into our analysis, we consider three different parameter scenarios:
$\alpha_{\rm c} \leqslant 0$, $0<\alpha_{\rm c}<Q^2 /8$, and $\alpha_{\rm c} =Q^2/8$. For all three scenarios, we further explore the coexistence volumes of small and large black hole phases during $P$-$V$ phase transitions, and systematically examine the divergent behavior of the normalized thermodynamic curvature scalar
$R_{\rm N}$.

For the first scenario ($\alpha_{\rm c} \leqslant 0$), the critical behavior of $R_{\rm N}$ revealed a universal critical exponent of $2$ along both small and large black hole curves, accompanied by a universal coefficient of $-1/8$. This coefficient aligns precisely with the numerical results obtained for a van der Waals fluid. In particular, at the critical point, the normalized thermodynamic curvature $R_{\rm N}$ diverges to negative infinity in both the small and the large phases. In the limit of
$\widetilde{T} \rightarrow 0$, both phases exhibit $R_{\rm N}$ converging to $1/2$, a behavior that reflects their shared asymptotic property at low temperatures. In contrast to a van der Waals fluid system, we observed that weak repulsive interactions prevail in charged AdS black holes with quantum anomalies at lower temperatures.

For the second scenario ($0<\alpha_{\rm c}<Q^2 /8$), we have further extended our analysis of the phase transition and critical behaviors of the quantum-anomalous charged AdS black hole, and additionally derived the coexistence volumes corresponding to the small and large black hole phases. Furthermore, our investigation of the critical behavior of the normalized scalar curvature $R_{\rm N}$ reveals two key universal features along both the coexistence curves of the small and large black hole phases: a universal critical exponent of $2$, and a universal coefficient of $-1/8$. This coefficient is in precise agreement with the numerical results reported for the van der Waals fluid. In particular, $R_{\rm N}$ exhibits divergent behavior in both phases, tending to as the system approaches the critical point. It is also worth emphasizing that for temperatures both below ($T<T_{\rm c}$) and above ($T>T_{\rm c}$) the critical temperature, $R_{\rm N}$ can take positive values when the temperature deviates from the critical point. This observation implies that, during the phase transition of quantum-anomalous charged AdS black holes, the microstructural interactions within the coexisting small and large black hole phases undergo a transition from attractive to repulsive.

For the third scenario ($\alpha_{\rm c} = Q^2 /8$), although $R_{\rm N}$ exhibits a universal critical exponent of $2$ along both the small and large black hole curves, its coefficient of $-2 \pm \sqrt{3}$ differs significantly from the corresponding value of $-1/8$ obtained for a van der Waals fluid. Moreover, at the critical point, the normalized thermodynamic curvature $R_{\rm N}$ diverges to negative infinity ($R_{\rm N}\rightarrow -\infty$) in both the small and large phases. It is noteworthy that for temperatures below the critical temperature ($T<T_{\rm c}$), $R_{\rm N}$ changes sign to positive in both the small and large black hole phases, suggesting that a weak repulsive interaction dominates in charged AdS black holes with quantum anomalies at low temperatures.
For temperatures above the critical temperature ($T>T_{\rm c}$), $R_{\rm N}$ remains negative along the coexistence curve for the large black hole phase, indicating that an attractive interaction dominates in charged AdS black holes with quantum anomalies. However, along the coexistence curve for the small black hole phase, $R_{\rm N}$ can be positive as the temperature deviates from the critical temperature, implying that the microstructural interaction in the coexistence small phase of charged AdS black holes with quantum anomalies undergoes a transition from attractive to repulsive during the phase transition.

The exploration of black hole thermodynamics, enriched by the incorporation of quantum conformal anomaly, has opened up fresh perspectives on the geometric understanding of black hole microstructures. By delving into the thermodynamic attributes and employing Ruppeiner geometry within the framework of a black hole model that incorporates quantum conformal anomaly, we aim to unravel the intricate degrees of freedom that fundamentally dictate black hole behavior. This endeavor holds profound implications for advancing our grasp of quantum gravity and the foundational fabric of spacetime.
It would be particularly intriguing to extend our research to a broader spectrum of black hole systems exhibiting diverse properties, particularly those where traditional scaling laws falter. Such an expansion would not only enrich our understanding but also provide a more nuanced view of the underlying degrees of freedom that underpin black hole thermodynamics. By doing so, we anticipate gaining substantially more detailed insights into the fundamental mechanisms governing these enigmatic objects, thereby further refining our comprehension of the quantum and gravitational realms.


\section*{Acknowledgements}

We thank the anonymous referee for his/her constructive comments and suggestions, which have greatly contributed to the improvement of this work.
This work was supported by the National Natural Science Foundation of China (NSFC) under grant No. 12465012, the Kashi University high-level talent research start-up fund project under grant No. 022024002, and Tianchi Talented Young Doctors Program of Xinjiang Uyghur Autonomous Region.

\appendix

\section{Temperature-Independent Thermodynamic Singularities For $0<\alpha_{\rm c} \leqslant {Q^2}/8$}\label{Appendix}

This appendix elaborates on the temperature-independent thermodynamic singularities arising in the parameter range $0<\alpha_{\rm c} \leqslant {Q^2}/8$, along with their interpretation in the $\widetilde{P}$-$\widetilde{V}$ phase diagram.

\begin{figure}[h]
\includegraphics[scale=0.7]{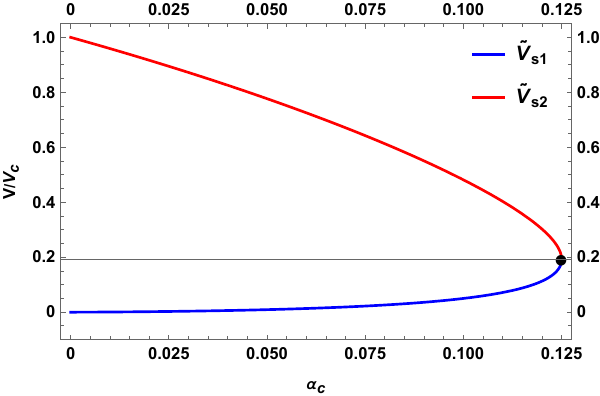}
\caption{Dependence of singular volumes $\widetilde{V}_{s1}$ and $\widetilde{V}_{s2}$ on $\alpha_{\rm c}$. The black dot denotes their intersection point with coordinates $(\alpha_{\rm c}, \widetilde{V}_{s})=(1/8, 1/3\sqrt{3})$, where the two volumes coincide. Here, we set $Q=1$.}\label{FigVs12}
\end{figure}

As indicated in the main text, when the first term in Eq.(\ref{EoSPVT1}) vanishes, we obtain the condition for temperature-independent thermodynamic singularities
\begin{equation}\label{V10}
\sqrt[3]{\pi}(6 V_{\rm c} \widetilde{V})^{2/3} -16 \pi \alpha_{\rm c} =0 \,.
\end{equation}
Substituting the critical volumes $V_{\rm c}$ from critical points (\ref{crit}) and (\ref{crit20}) into Eq.(\ref{V10}) respectively, the corresponding reduced volumes associated with these thermodynamic singularities are derived as
\begin{equation}\label{V12}
\widetilde{V}_{s1}=\frac{8 \alpha_{\rm c}^{3/2}}{\sqrt{\left(3 Q^2+K-12\alpha_{\rm c}\right)^3}} \,,\,\,\widetilde{V}_{s2}=\frac{8\alpha_{\rm c}^{3/2}}{\sqrt{\left(3 Q^2-K-12\alpha_{\rm c}\right)^3}}\,. \nonumber
\end{equation}
To intuitively illustrate the dependence of these singular volumes on the parameter $0<\alpha_{\rm c} \leqslant Q^2 /8$, Fig.\ref{FigVs12} is provided below.

It is evident from the analysis that within the interval $0<\alpha_{\rm c} < Q^2 /8$, each value of $\alpha_{\rm c}$ corresponds to two distinct singular volumes $\widetilde{V}_{s1}$ and $\widetilde{V}_{s2}$. This implies the existence of two separate temperature-independent thermodynamic singularities in the $\widetilde{P}$-$\widetilde{V}$ phase diagram for any $\alpha_{\rm c}$ in this range. These singularities manifest as special points where thermodynamic quantities that are not governed by temperature, a key characteristic distinguishing them from temperature-dependent phase transition points.
When $\alpha_{\rm c}$ reaches the upper bound $\alpha_{\rm c} = Q^2 /8$, the two singular volumes $\widetilde{V}_{s1}$ and $\widetilde{V}_{s2}$ coincide. This convergence implies the merging of the two thermodynamic singularities into a single singular point, with the corresponding singular volume being $\widetilde{V}_{s}=1/3\sqrt{3}$.

In the $\widetilde{P}$-$\widetilde{V}$ phase diagram, these temperature-independent singularities correspond to vertical or otherwise distinct features that are insensitive to temperature variations. Unlike conventional phase transition curves (which shift with temperature), these singular points maintain fixed positions in the $\widetilde{P}$-$\widetilde{V}$ plane for a given $\alpha_{\rm c}$, reflecting their temperature-independence. Their presence provide crucial insights into the intrinsic thermodynamic stability of the system under the constraint $0<\alpha_{\rm c} < Q^2 /8$.

\end{document}